\documentstyle{mn}
\def\be{\begin{equation}}
\def\ee{\end{equation}}
\def\intii{\int_{-\infty}^{+\infty}}
\def\vpe{v_{\perp}}

\clearpage
\newpage

\title[Dynamics of thin disks]{From kinematics to dynamics in thin
galactic disks}

\author[A.~Mathieu and M.R.~Merrifield]{A.~Mathieu and
M.R.~Merrifield\\ University of Nottingham, School of Physics and
Astronomy, University Park, Nottingham NG7 2RD}
\date{Received date; accepted date}

\begin{document}
\label{firstpage}
\maketitle

\begin{abstract}

We present a method for recovering the distribution functions of
edge-on thin axisymmetric disks directly from their observable
kinematic properties.  The most generally observable properties of
such a stellar system are the line-of-sight velocity distributions of
the stars at different projected radii along the galaxy.  If the
gravitational potential is known, then the general two-integral
distribution function can be reconstructed using the shapes of the
high-velocity tails of these line-of-sight distributions.  If the
wrong gravitational potential is adopted, then a distribution function
can still be constructed using this technique, but the low-velocity
parts of the observed velocity distributions will not be reproduced by
the derived dynamical model.  Thus, the gravitational potential is
also tightly constrained by the observed kinematics.

\end{abstract}


\section{Introduction}
The dynamics of a collisionless stellar system is fully specified by
its distribution function (DF) -- the phase density of its constituent
stars -- and the form of the gravitational potential that binds it
together.  The most fundamental goal of galactic dynamics is the
recovery of this information from the observable properties of a
galaxy.  There is some hope that this objective might soon be
achieved: sophisticated data analysis techniques applied to the
Doppler broadening of galaxy spectra now routinely yield reasonable
estimates for the full distribution of the stars' line-of-sight
velocities, generally referred to as their line-of-sight velocity
distribution, or LOSVD (see e.g. Franx \& Illingworth 1988, Bender
1990, Rix \& White 1992, van der Marel \& Franx 1993, Winsall \&
Freeman 1993, Kuijken \& Merrifield 1993).  In principle, one can
therefore measure a complete three-dimensional function, the density
of stars as a function of both their line-of-sight velocities and the
two spatial coordinates giving the position on the plane of the sky.
Since the DF of a stellar system in equilibrium depends on at most
three isolating integrals of motion (Binney \& Tremaine 1987), it is
plausible that one might deproject the observed three-dimensional
function into the galaxy's intrinsic DF.

Although this dimensional argument suggests that the inversion might
be carried out, we have no guarantee that the solution will be unique:
it is possible that more than one DF may produce identical LOSVDs.
Further, the dynamics of a galaxy depends on the form of the
gravitational potential that confines it.  It is therefore also
possible that a given set of kinematic observations will be consistent
with different dynamical models depending on the form adopted for the
gravitational potential.  Thus, we do not yet know whether the
dynamics of a galaxy is uniquely specified by its observable
properties.

In some special cases, we can make more definitive statements.  For a
spherical system with a known gravitational potential, for example, it
has been proved that the dynamics is completely specified by the
observable kinematics.  Dejonghe \& Merritt (1992) considered the
moments of the velocity distribution -- the velocity dispersion and
its higher order analogues.  For each order, they demonstrated that
the intrinsic moments of a spherical galaxy's velocity distribution
could be inferred from the observable line-of-sight component.  They
thus elegantly proved that all the intrinsic dynamical properties of
the galaxy are specified by observable quantities.  Dejonghe \&
Merritt also showed that the observable kinematics will limit the
possible forms of the gravitational potentials that might be confining
the system, although not necessarily to the point of specifying it
uniquely.

A further significant advance was made by Merritt (1996), who
considered edge-on axisymmetric systems where the DF respects two
integrals of motion.  He showed that for such systems one can use the
information in the projected density of stars and the first two
moments of their line-of-sight velocity distributions to estimate
both the DF and the gravitational potential.  Although not
constituting a formal proof of uniqueness, the numerical solutions to
Merritt's equations do seem to indicate that accurate dynamical models
can be inferred from the observable kinematic properties.

In theoretical studies of galaxy dynamics, manipulation of the moments
of the velocity distribution is mathematically attractive because
these quantities obey the Jeans equations and their higher-order
analogues (Merrifield \& Kent 1990, Dejonghe \& Merritt 1992).
However, the application of such analyses to observations is, in
practice, fraught with difficulties.  The higher-order moments on
which Dejonghe \& Merritt's (1992) proof depend are extremely
sensitive to noise, so cannot be calculated from real data.  Even
the velocity dispersion, calculated from the second moment of the
LOSVD, is difficult to estimate from noisy data, particularly if the
velocity distribution has long tails out to high velocities.  In
poorer-quality data, the velocity dispersion is often estimated by
fitting a Gaussian to the velocity distribution; there is in reality
little justification for such a procedure, and systematically-biassed
estimates for the dispersion are almost certain to be the result.

In this paper, we therefore adopt a somewhat different approach.  The
prevalence of techniques for measuring full LOSVDs suggests that it
might be worth moving away from analysis of moments to using the
complete shape of the velocity distribution directly in the analysis.
To this end, we have been investigating the properties of a thin
axisymme\-tric disk model.  In Section~2, we discuss how the dynamics of
an inclined thin disk galaxy can be inferred directly from its
observable kinematics.  Such a simplified model is clearly a rather
special case, since each line of sight passes through the disk at only
a single point.  Therefore, in Section~3 we consider the extra
complexity that is introduced when such a disk is viewed edge-on, so
that each line of sight passes through many points in the galaxy.  We
have used the ana\-ly\-sis of the inclined case to provide the basis for
an iterative scheme for recovering the DF of such an edge-on disk.  In
Section~4, we test this algorithm on a realistic disk model, and show
how the DF can, indeed, be recovered from the observable kinematics.
We further show that even small errors in the assumed gravitational
potential lead to inconsistencies in the inferred kinematic
properties, and hence that the potential is very tightly constrained
by the observable properties of such a galaxy.  Section~5 discusses
the possible practical application of this analysis to studies of disk
galaxies.

\section{Kinematics of thin disks}

The simplest plausible model for a disk galaxy is one in which the
system is axisymmetric, with all the motion confined to a single
plane.  For such a system, Jeans' theorem states that, in equilibrium,
the DF can depend only on two isolating integrals of motion, the
energy of stars in the disk plane, and the angular momentum of stars
about the centre of the galaxy (Binney \& Tremaine 1987).  Defining
the usual polar coordinates about the centre of the system and wri\-ting
the gravitational potential in the plane of the galaxy as $\Psi(r)$,
these integrals can be written as $E = \frac{1}{2}(v^2_r + v^2_\phi) +
\Psi(r)$, and $L = r\, v_\phi$, and the DF can be expressed as a
function $f(E,L)$.

At any radius in the galaxy, the tangential velocity distribution can
be obtained by integrating the DF over the whole range of radial
velocity:
\be
\tilde f_\phi(r,v_\phi) = \intii f(E,L)\, dv_r = 2 
\int_{W_\phi}^{\infty}
\frac{ f(E,L)\, dE}{\sqrt{2(E-W_\phi)}},
\label{abeleq}
\ee
where 
\be
W_\phi = \Psi(r) + \frac{v^2_\phi}{2}.
\ee
If we transform variables and write 
\be
\tilde f_\phi(r,v_\phi) \equiv f_\phi(W_\phi,L),
\ee
then equation~(\ref{abeleq}) is an Abel integral equation relating
$f(E,L)$ and $f_\phi(W_\phi, L)$.  Thus, if $f_\phi$ is known, one can
use a standard Abel inversion to solve for the full DF, 
\be
f(E,L) = -\frac{1}{\pi} \int_{E}^{\infty}
                   \frac{ \partial f_\phi(W_\phi,L)}{\partial W_\phi}
                   \frac{ dW_\phi}{\sqrt{2(W_\phi-E)}}.
\label{abelinv}
\ee

For a thin disk at inclination $i$ to the line of sight, a quantity
related to $f_\phi$ that one can observe is the line-of-sight velocity
distribution as a function of projected radius along the galaxy's
apparent major axis, $F_{maj}(r_p, v_{los})$.  Via some simple
geometry, one can show that
\be
F_{maj}(r_p, v_{los}) \equiv \tilde f_\phi(r = r_p, v_\phi = 
v_{los}/\sin
i)/\sin i.
\label{Fmaj}
\ee
Thus, as first pointed out by Merrifield \& Kuijken (1994), if one
observes the LOSVD along the major axis of such a disk, then one can
readily derive $f_\phi$ and hence solve for the DF by inverting
equation~(\ref{abeleq}).  More recently, a sophisticated investigation
of this inversion has been made by Pichon \& Thi\'ebaut (1998), who
showed that it is possible to regularize a non-parametric algorithm to
carry out the requisite Abel inversion without excessive noise
amplification.

There is a degree of redundancy in the inversion, which means that its
implementation can be simplified somewhat further.  Every pair of
$\{W_\phi, L\}$ values corresponds to two pairs of $\{r, v_\phi\}$
values, and hence two points in the obser\-vable $\{r_p, v_{los}\}$
parameter space.  Thus, $f_\phi (W_\phi, L)$ is over\-spe\-ci\-fied by the
observed $F_{maj}(r_p, v_{los})$.  If we define the galaxy's circular
speed, $v_c(r)$, by the usual relation, \be v_c^2(r) = r {d\Psi \over
d r}, \ee then it is straightforward to show that the DF is completely
specified by both the part of $F_{maj}(r_p, v_{los})$ where $|v_{los}|
> v_c(r_p)\sin i$ and by the part where $|v_{los}| < v_c(r_p)\sin i$
(Merrifield \& Kuijken 1994).

Using just one part of the LOSVD allows us to simplify the
quadrature in equation~(\ref{abelinv}).  For the high-velocity
side of the LOSVD in a realistic disk, $f_\phi(W_\phi,L)$ is a
monotonically decreasing function of $W_\phi$ at a given value of the
angular momentum $L$. 
One can show that any positive distribution function such that
\be
\frac{\partial f(E,L)}{\partial E} < 0 
\label{dfeneg}
\ee
yields a tangential velocity distribution $f_\phi(W_\phi,L)$ with the
above property.
In stellar dynamics one usually considers systems with distribution
functions that obey equation~(\ref{dfeneg}). Definite criteria for the
stability of spherical and axisymmetric stellar systems have been
established for distribution functions which are decreasing functions
of the energy (Antonov 1962).  Moreover it is found that most
distribution functions used to produce realistic models of galaxies do
actually obey that condition. On the other hand, very little work,
either numerical or theoretical, has been done using distribution
functions which do not satisfy such a condition and very little is
known about the stability of these systems (H\'enon 1973, Perez \& Aly
1996). Therefore, in our paper, we follow the generally accepted
assumption given by equation~(\ref{dfeneg}).
In this case  $W_\phi$ can be written as a monotonic
function of $f_\phi(W_\phi,L)$ at given $L$.  This property allows us
to transform the above relation into
\be
f(E,L) = \frac{1}{\pi} \int_{0}^{f_\phi(W_\phi = E,L)}
\frac{ d{f}'_\phi}{\sqrt{2(W_\phi({f}'_\phi,L))-E)}}\,.
\label{simpleabel}
\ee
Provided that the function $f_\phi(W_\phi,L)$ can be numerically
inverted to give $W_\phi(f_\phi,L)$, this equation means that the DF
can be calculated with a simple one-dimensional integration, bypassing
the computationally-unstable estimation of the derivative of
$f_\phi(W_\phi,L)$ in equation~(\ref{abelinv}).

\section{Kinematics of edge-on thin disks}
\label{edgeonsec}
As mentioned in the Introduction, an inclined thin disk is rather
simpler than most realistic galaxy models, since any line of sight
will only intersect it at a single point.  Further, for an inclined
system, one cannot tell from the photometry whether the assumption
that the disk is thin is a valid one -- it is only when the system is
viewed edge-on that its razor-thinness will be unambiguously apparent.
We therefore now turn to consider the case of such a system viewed
edge-on.  In this case, the LOSVD is not just a rescaled version of
$f_\phi(W_\phi,L)$ as in equation~(\ref{Fmaj}), but instead is
generated by an integration along the line of sight through the entire
disk. By choosing the $z$-axis as the direction of the line of sight,
one can write the LOSVD at some distance $r_p$ from the centre of the
galaxy as
\be 
F(r_p,v_{los}) = \intii dz \intii d \vpe\, f(E,L)
\label{projop}
\ee 
where $(v_{los},\vpe)$ are the components of the velocity parallel and
perpendicular to the direction of the line of sight respectively.  The
two components of the velocity $(v_r,v_\phi)$ can be written in terms
of $(v_{los},\vpe)$ as:
\begin{eqnarray}
v_r &=& \vpe \cos \phi  + v_{los} \sin \phi;\\
v_\phi &=& v_{los} \cos \phi  - \vpe \sin \phi,
\end{eqnarray}
with
\be 
\cos \phi = \frac{r_p}{r}; \qquad \sin \phi = \frac{z}{r}; 
\qquad r =\sqrt{r^2_p + z^2}.
\ee
The integrals of motion are then
\begin{eqnarray}
E &=& \frac{1}{2} ( \vpe^2 + v_{los}^2) + \Psi(r)\\
L &=& r\, v_\phi = r_p\, v_{los} - z\, \vpe.
\end{eqnarray}

The extra integration along the line of sight clearly complicates the
relationship between the observed kinematics and the intrinsic
dynamics of the galaxy.  However, the properties that we have derived
above for inclined disks suggest a way to proceed.  In particular, the
fact that the DF is completely specified by the high-velocity tail of
$\tilde f_\phi$ can be used to our advantage.  In general, the
mean-streaming circular motions of the stars in a realistic disk will
be significantly greater than their random motions, so one can think
of stars at each radius moving with a mean streaming velocity of close
to the local circular speed, with a smaller amount of random motion
superimposed.  Thus, the LOSVD at a projected radius $r_p$ in an
edge-on disk will be peaked at a velocity close to $v_c(r_p)$.  At
lower line-of-sight velocities, there will be contributions both from
stars that have intrinsically low velocities and from stars that lie
at large radii in the galaxy, so that their large circular motions are
oriented mostly transverse to the line of sight, resulting in a small
line-of-sight component.  However, at line-of-sight velocities greater
than $v_c(r_p)$, the majority of the stars contributing to the LOSVD
will be those where the circular motion is oriented along the
line-of-sight, which occurs only for stars at radii $r \sim r_p$.
Further, for the stars at radii close to $r_p$, the line-of-sight
component of their random motions measures their velocities in the
$\phi$ direction.  Thus, to a simple approximation, for $v_{los} >
v_c$, 
\be
F(r_p, v_{los}) \approx \tilde f_\phi(r = r_p, v_\phi = v_{los})
\times \Lambda, 
\label{Lambdadef}
\ee 
where $\Lambda$ is the length of path through the galaxy that lies
sufficiently close to $r = r_p$ for there to be a significant
contribution to $F(r_p, v_{los})$.  In terms used in the study of the
Milky Way, this path-length covers the region of the tangent point,
over which $\tilde f_\phi$ will not change rapidly (since $r$ is not changing
significantly), so the integral along the line of sight can be
replaced by the simple product of equation~(\ref{Lambdadef}).
Thus, equation~(\ref{Lambdadef}) implies that by observing the part of
the LOSVD at high velocities, we obtain an estimate for the shape of
the high-velocity part of the tangential velocity distribution, which
we need to estimate the DF.

The close relation between $F(r_p, v_{los})$ and $\tilde f_\phi(r,
v_\phi)$ provides a good indication that the inversion of
equation~(\ref{projop}) may not be significantly more ill-conditioned
than was the case for the simpler case of the inclined disk
[equation~(\ref{abeleq})].  Although equation~(\ref{projop}) contains
an extra integration, which often serves to smooth out some of the
information in the integrand, in this case much the same information
is contained in the double-integral [$F(r_p, v_{los})$] as there was
in the simpler single-integral [$\tilde f_\phi(r, v_\phi)$].  This
discovery bodes well for the solution of equation~(\ref{projop}),
since, as Pichon \& Thi\'ebaut (1998) have ably demonstrated, it is
quite possible to regularize the single-integral case to a point where
it is realistically soluble.

If we are in practice to convert $F(r_p, v_{los})$ into $\tilde
f_\phi(r, v_\phi)$, we need the value for the normalization factor,
$\Lambda$ in equation~(\ref{Lambdadef}).  On the basis of simple
geometric arguments, one might expect $\Lambda$ to be proportional to
$r_p$.  However, the exact value $\Lambda$ cannot be determined in
any simple way.  Indeed, one would expect the value to depend on the
DF of the disk, the very thing that we are trying to derive.  However,
as we shall see below, the similarity between the shapes of $F(r_p,
v_{los})$ and $\tilde f_\phi(r_p, v_\phi)$ does form the basis for a
useful iterative scheme for recovering the DF, which is completely
insensitive to the form adopted for $\Lambda$.

In order to describe this scheme, we define some simplified notation.
Let ${\cal A}^{-1}$ be the Abel inversion operator of
equation~(\ref{abeleq}) and ${\cal P}$ the line-of-sight
projection operator in equation~(\ref{projop}). In this notation, the
line-of-sight velocity distribution, the DF of the system and the
tangential velocity distribution are related as follows:
\be
F(r_p,v_{los}) = {\cal P} f(E,L) = {\cal P}({\cal A}^{-1} \tilde f_\phi).
\label{opdef}
\ee
We can now present an iterative algorithm which computes velocity
distributions that converge towards the true tangential velocity
distribution.  As discussed above, the part of the tangential velocity
distribution above the rotation curve is sufficient to uniquely
specify the DF, and so for the remainder of this section we consider
only this part of the distribution.  Let $\hat{f}_\phi^{(n)}$ be the
${n}^{\rm th}$ approximation to the tangential velocity distribution.
As discussed in the previous paragraph, $\hat{f}_\phi^{(0)} =
\Lambda^{-1} F$ would be a good zeroth-order estimate if we knew the
value for $\Lambda$.  However, even if $\Lambda$ is known, this
estimate is only an approximation, so ${\cal A}^{-1}
\hat{f}_\phi^{(0)}$ will differ somewhat from $f(E,L)$ and hence
${\cal P}({\cal A}^{-1}(\hat{f}_\phi^{(0)}))$ will have systematic
residuals from the observed form of $F$ that we started with.  As a
next iteration, we correct for these small residuals by subtracting
them from our assumed form for the tangential velocity distribution,
so that
\be
\hat{f}_\phi^{(n+1)} = \hat{f}_\phi^{(n)} - \Lambda^{-1} \left[{\cal
P}({\cal A}^{-1} ( \hat{f}_\phi^{(n)})) - F\right],
\label{iteration}
\ee 
and iterate until convergence is achieved.  If this scheme converges,
we know that $\hat{f}_\phi^{(n+1)} \rightarrow \hat{f}_\phi^{(n)}$.
It therefore follows from equation~(\ref{iteration}) that 
\be
{\cal P}({\cal A}^{-1} ( \hat{f}_\phi^{(n)})) \rightarrow F.  
\label{convergence}
\ee 

Note that, with this formulation, if the iteration converges then the
result will be totally independent of the adopted form for $\Lambda$,
so our ignorance as to the exact value for this parameter is not a
problem -- $\Lambda$ acts as a relaxation parameter in the iteration,
which may affect the speed of convergence, but not the final answer.
Thus, as the iteration converges, from equation~(\ref{opdef}), we have
$\hat{f}_\phi^{(n)} \rightarrow \tilde f_\phi$.  One can then obtain
the distribution function by a simple Abel inversion, $f(E,L) = {\cal
A}^{-1} \tilde f_\phi$.

To implement the Abel inversion operation, ${\cal A}^{-1}$, one can draw
on the sophisticated techniques developed by Pichon \& Thi\'ebaut
(1998) in their analysis of non-edge-on disks.  However, this paper is
primarily concerned with esta\-bli\-shing whether the iterative scheme
described above converges.  For this purpose, we can consider model data
in which the noise level is low, and apply the computationally-simpler
approach given by equation~(\ref{simpleabel}).  Numerically, at each
step, interpolations of the estimate of the DF, its projected velocity
distribution and the approximation of the tangential velocity
distribution $\hat{f}_\phi^{(n)}$ are computed. We constrain the
tangential velocity distribution and distribution function estimates
to be positive.  A satisfactory approximation of the DF is found when
the difference between the LOSVD $F$ and the projection of the
approximate DF is small everywhere in a large region of the
$(r_p,v_{los})$ plane.  The resolution of the interpolation grid
ultimately limits the accuracy with which the DF can be recovered with
this iterative scheme, eventually resulting in noise amplification.

\section{Model tests of the algorithm}

To test the above proposed algorithm, we have constructed the
line-of-sight velocity distribution as a function of projected radius
for a moderately-realistic edge-on disk model.  The DF was taken to be
of the form
\be 
f = g(L)\exp \left[ -\beta (L) (E - E_c(L)) \right],
\label{dfdef}
\ee 
where $g(L)$ and $\beta(L)$ are two arbitrary functions of the
angular momentum and $E_c(L)$ is the energy of the circular orbit of
angular momentum $L$. Similar forms of DFs were described in previous
work by Binney (1987) and Kuijken \& Tremaine (1991).  For these
tests, we adopt functional forms of 
\be 
\beta(L) = \exp \left(\frac{L}{L_0} \right)
\label{betadef}
\ee
and 
\be
g(L) = \left\{
\begin{array}{ll}
\exp \left( -\frac{L}{L_0}\right) \quad &{\rm if \ } L \geq 0\\
0 \quad &{\rm otherwise.}
\end{array}
\right.
\label{gdef}
\ee
To complete the model, we must also specify the gravitational
potential.  In order to match the flat rotation curves found in real
galaxies, we adopt the softened isothermal sphere potential, 
\be 
\Psi(r) = \frac{v^2_0}{2} \ln \left( 1 + \frac{r^2}{r^2_0} \right)
\label{Psidef}
\ee

\begin{figure}
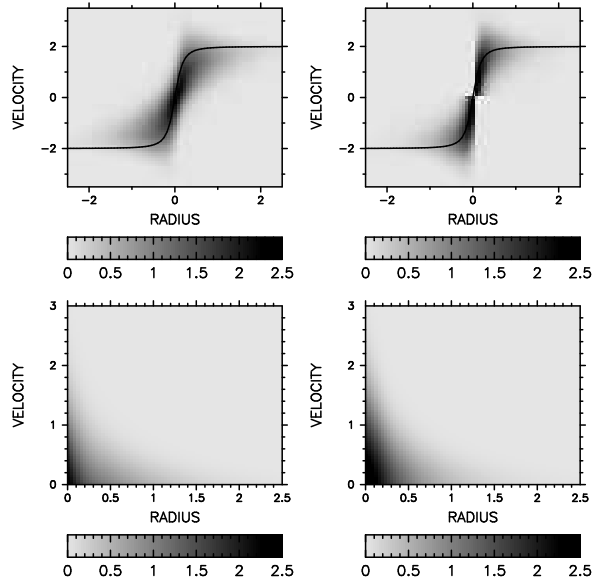

\vspace*{8.5cm}
\special{hscale=40 vscale=40 hoffset=-10 voffset=0
hsize=500 vsize=800 angle=0 psfile="fig1.ps"}
\caption{Top: line-of-sight velocity distribution $F(r_p,v_{los})$
(left) and tangential velocity distribution (right)
$\tilde f_\phi(r,v_\phi)$; the line shows the adopted rotation
curve. Bottom: plot of $F(r_p,v_{los} + v_c(r_p))$ and
$\tilde f_\phi(r,v_\phi + v_c(r))$   for the region above the rotation
curve.}  
\label{losvd}
\end{figure}

The upper panels of Fig.~\ref{losvd} show the LOSVD as a function of
projected radius and the tangential velocity distribution as a
function of radius for this model.  The model does, indeed, look very
similar to the LOSVDs seen in the stellar kinematics of real edge-on
disk galaxies (e.g. Kuijken, Fisher \& Merrifield 1996).  The lower
panels show the parts of these two functions that lie ``above'' the
rotation curve.  For the reasons discussed in Section~\ref{edgeonsec},
these two plots appear similar, allowing us to use the bottom
right panel as our initial approximation for the bottom left panel.

\begin{figure*}
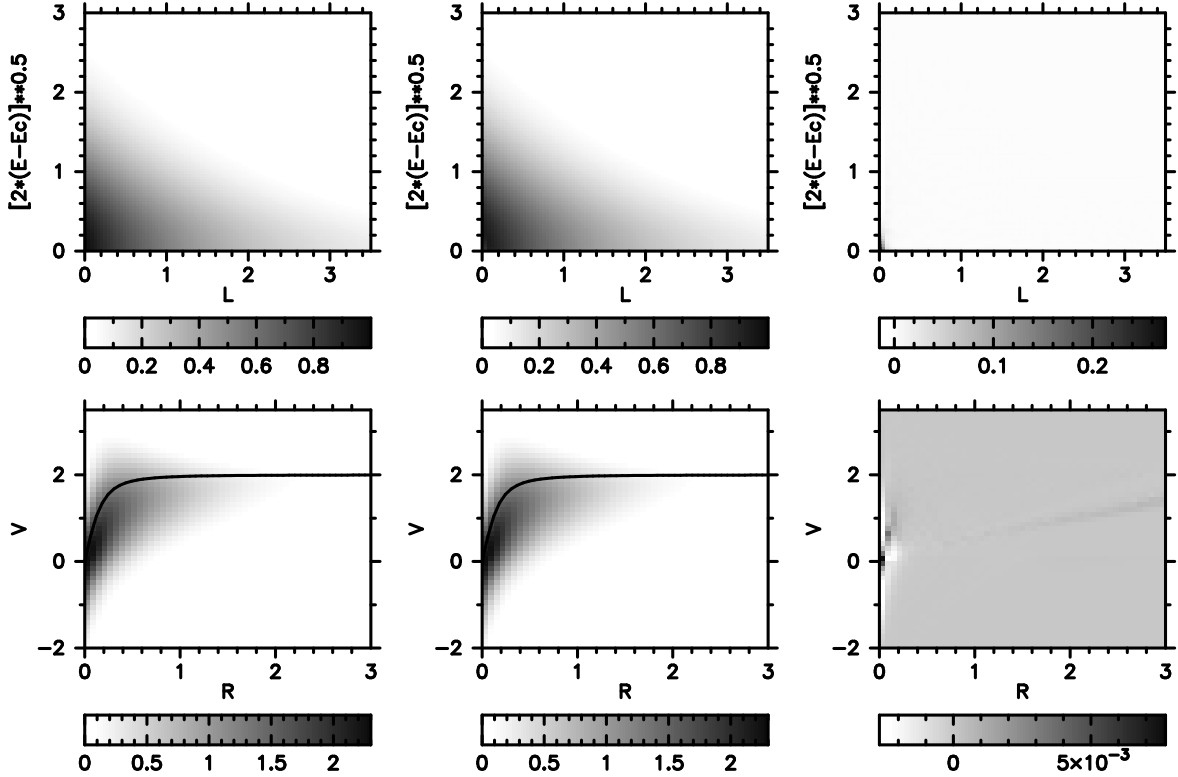

\vspace*{12.5cm}
\special{hscale=80 vscale=80 hoffset=0 voffset=-130
hsize=500 vsize=800 angle=0 psfile="fig2.ps"}
\caption{
The distribution function and LOSVD for the simple dynamical model described 
in the text, as recovered by the iterative algorithm  Top: the true 
distribution function expressed as $f(\sqrt{2[E - E_c(L)]},L)$ (left), the 
distribution function recovered by the algorithm (middle), and absolute 
difference between the two (right).  Bottom: the true LOSVD (left), the LOSVD 
recovered by the algorithm (middle), and absolute difference between the two 
(right).}
\label{approxlosvd}
\end{figure*}

We have applied the algorithm described in Section~\ref{edgeonsec} to
a model of the form given by Eqs.~(\ref{dfdef}), (\ref{betadef}),
(\ref{gdef}) and (\ref{Psidef}) for a variety of sets of parameters.
Figure~\ref{approxlosvd} illustrates the case for parameters $L_0 =
1.6$, $r_0=0.2$, $v_0=2.0$. As this figure shows, the iterative
algorithm recovers the DF very closely.  In fact, convergence occurs
in a small number of steps.  The only remaining discrepancies between
the ``observed'' LOSVDs and those produced by projecting the recovered
DF occur at very small values of $r_p$, due to numerical noise in the
integration process.

Thus far, we have assumed that we know the gravitational potential in
our reconstruction of the DF.  Such an assumption is reasonable if
modelling a disk galaxy containing gas from which an emission-line
rotation curve can be obtained.  However, such information would not
be available for a purely stellar system such as an S0 galaxy.  
Further, we are also interested in addressing the more general
question of whether the gravitational potential is uniquely specified
by the observable stellar kinematics, or whether one can derive
equally-plausible distribution functions using different assumptions
about the form of the potential.  

It is apparent from Fig.~\ref{losvd} that there is no obvious way to
estimate the rotation curve from the observed LOSVDs: the combination
of asymmetric drift in the stellar kinematics and the effects of
projection along the line of sight means that the local circular speed
does not correspond to any simple property of the stellar kinematics
such as the peak of the LOSVD or the mean line-of-sight velocity of
the stars.  We are therefore, in principle, free to choose a different
rotation curve and hence gravitational potential.

\begin{figure*}
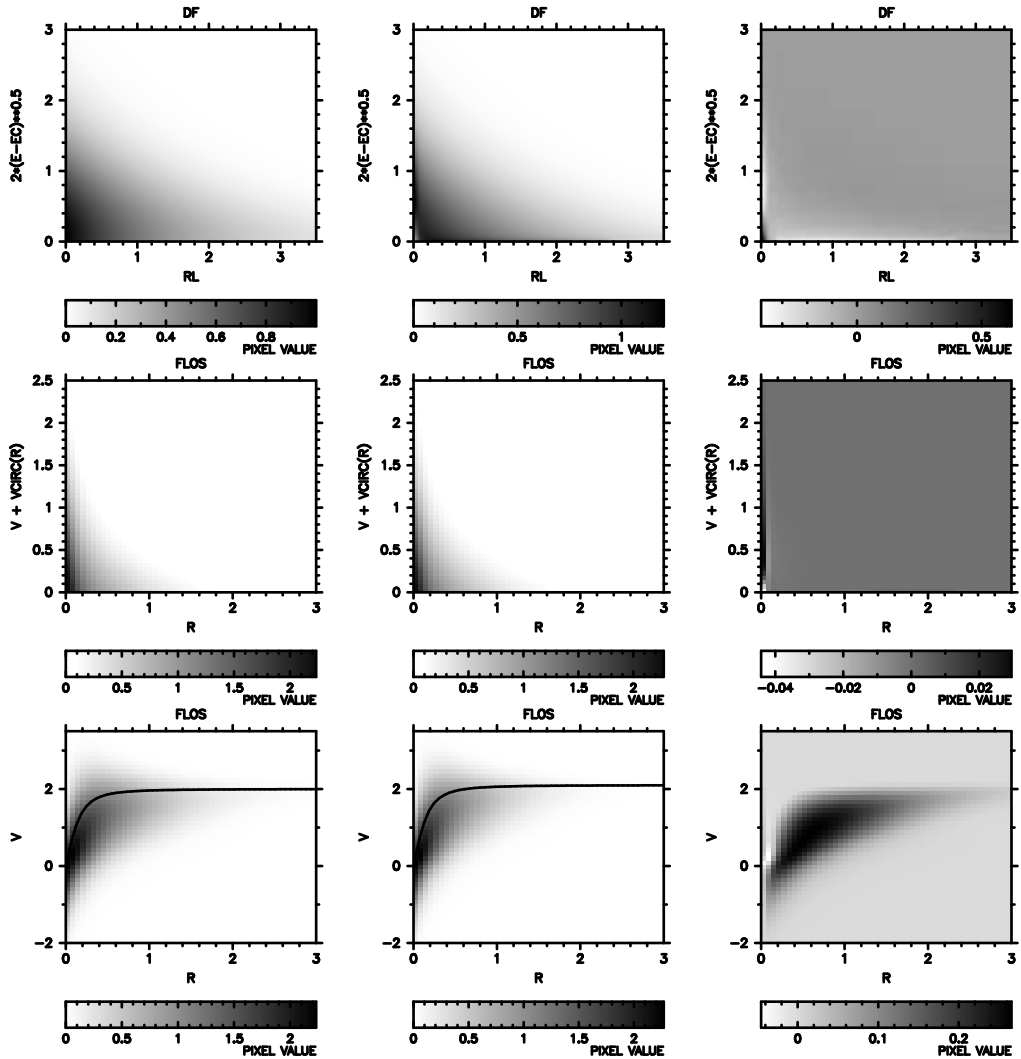

\vspace*{14.5cm}
\special{hscale=70 vscale=70 hoffset=50 voffset=-140
hsize=500 vsize=500 angle=0 psfile="fig3.ps"}
\caption{Illustration of the effects of assuming a slightly erroneous
rotation curve.  The left panels show the true properties of the model,
the middle panels shows the dynamical properties recovered by the
iterative algorithm with the wrong adopted potential, and the right
panels show the difference between the two.  The top row shows the
distribution functions, the middle panel shows the high-velocity part
of the LOSVD, and the bottom panels show the full LOSVDs.  The
parameters of the true potential are $(r_0=0.2,v_0=2.0)$ and the
parameters used for the reconstruction are $(r_0=0.2,v_0=2.1)$ [see
equation~(\ref{Psidef})].
}
\label{v2.1}
\end{figure*}

\begin{figure*}
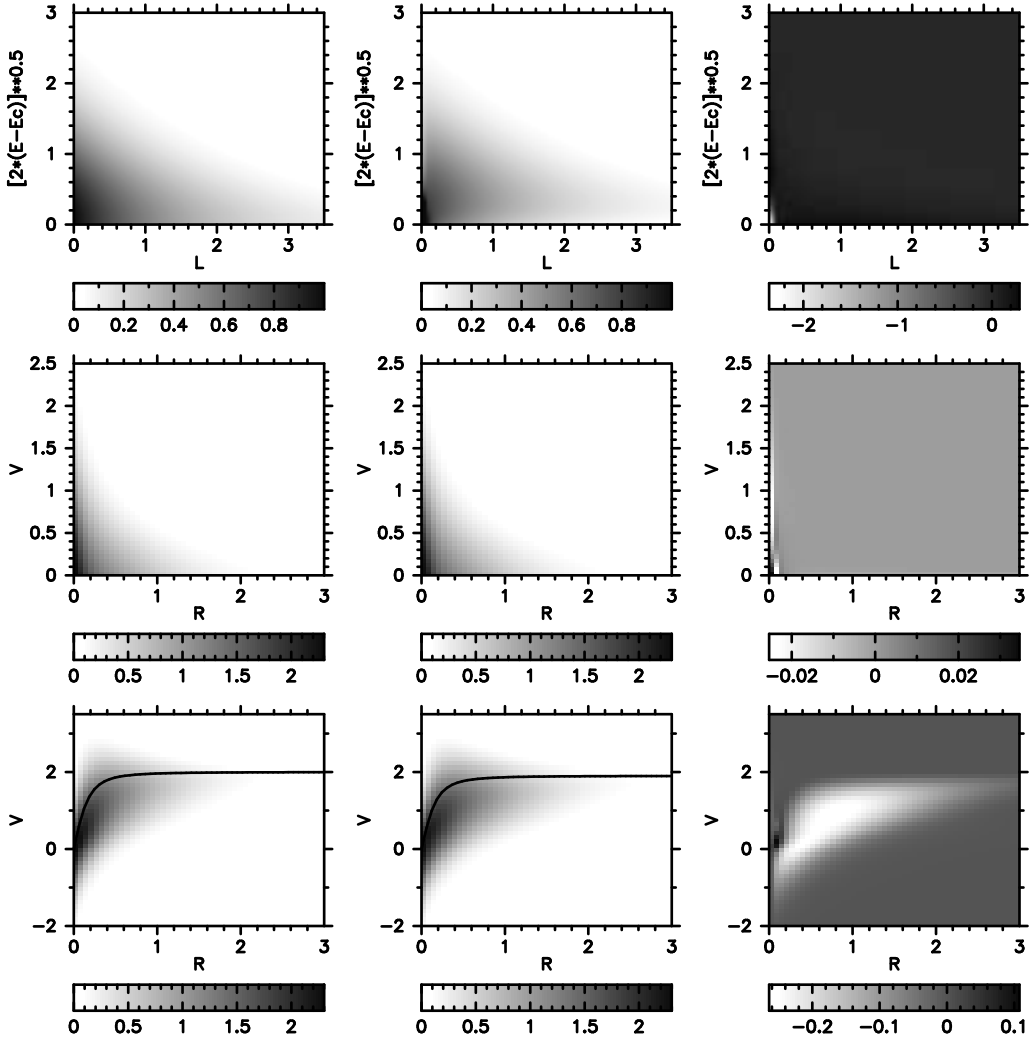

\vspace*{14.5cm}
\special{hscale=70 vscale=70 hoffset=50 voffset=-140
hsize=500 vsize=500 angle=0 psfile="fig4.ps"}
\caption{Same as in Fig.~\ref{v2.1} except that now the parameters 
used for the reconstruction are $(r_0=0.2,v_0=1.9)$.
}
\label{v1.9}
\end{figure*}

Figures~\ref{v2.1} and~\ref{v1.9} show what happens in practice if we
do so.  For these model calculations, we have adopted a gra\-vi\-ta\-tio\-nal
potential that differs only fairly marginally from the true form.  The
iterative process again converges rapidly to a plausible DF, which
reproduces the LOSVDs exactly for $|v_{los}| > v_c(r_p)$.  However,
the LOSVDs that one predicts from the derived DF for $|v_{los}| <
v_c(r_p)$ bear little resemblance to those of the original galaxy
model.  Thus, it would appear that the exact form of the gravitational
potential is tightly constrained by the observations: using just the
high-velocity kinematics, one can reconstruct the full DF consistent
with any given gravitational potential, but the low-velocity tails of
the LOSVDs will only be correctly reproduced if the correct potential
is adopted.

\section{Conclusions}

The ultimate goal of dynamical astronomy is the derivation of all
that there is to know about a galaxy's dynamics from its observable
kinematics.  We are still clearly a long way from attaining this
``holy grail,'' but the analysis of this paper does provide some cause
for optimism.  Specifically, we have shown how the distribution
function of a relatively simple model galaxy can be estimated directly
from its observed kinematics, using a straightforward iterative scheme.
Further, the redundancy of information in the kinematics means that
one can readily rule out models in which the wrong gravitational
potential has been adopted.

The success of this iterative scheme seems to derive from the fact
that the information available from the LOSVD of an edge-on disk is
very similar to that available for inclined disks, for which the
inversion to the DF is already well established (Merrifield \& Kuijken
1994, Pichon \& Thi\'ebaut 1998).  Thus, perhaps rather surprisingly,
this problem does not appear significantly more ill-conditioned than
the simpler case of the inclined disk, even though an extra integral
is involved.

The most tempting practical application for this technique is the
study of edge-on S0 galaxies, since such objects are fairly pure
stellar disks, which are believed to contain little by way of
obscuration by dust.\footnote{It is worth mentioning that obscuration
will affect the high-velocity and low-velocity tails of the LOSVDs in
different ways, so in a galaxy where emission lines allow us to
determine the potential unambiguously, a mismatch between the model
and observed low-velocity tails of the LOSVD could be used as a
measure of extinction.}  There are, however, still several obstacles
to such an analysis.  For a start, S0 galaxies also contain a central
spheroidal bulge component, so any attempt to reproduce their
kinematics must include a suitable dynamical bulge model.  In
addition, even the disk components of these systems are not infinitely
thin, so they probably also obey a third integral of motion.  The
simplest models would treat the dependence on this third integral as a
separable function, but there is no reason why real galaxies should
follow such simple models.  Finally, the data that one obtains for
real galaxies will be noisy.  The derivation of LOSVDs from the
broadening of spectral lines is a fundamentally noise amplifying
process, so even good quality data will contain significant noise
contribution.  Further, in the case of the algorithm presented in this
paper, we principally use only the data from the high velocity side of
the LOSVD, which will not span a large range in velocities.  We will
therefore have to obtain rather high dispersion spectra to determine
this function, resulting in a further cost in terms of noise.
However, the advent of 8-metre class telescopes means that the high
quality data required for such analyses should soon be routinely
available. Also in its favour, the method works directly with the
observed LOSVD rather than the moments calculated from it.  Even the
second moment of the velocity distribution is very sensitive to noise,
so a method that avoids calculating such moments is likely to be
relatively robust.  We can also draw on sophisticated Abel inversion
techniques, such as those developed by Pichon \& Thi\'ebaut (1998), to
minimize any noise amplification.  

Clearly, the next step in this project should be to apply these
techniques to real data.  Preliminary DF reconstructions based on
LOSVDs obtained from observed spectra of edge-on S0 galaxies indicate
that the algorithm presented in this paper can be applied to real data
with attainable noise levels to yield an estimate of the underlying
DF; we defer discussion of this development to a subsequent paper
(Mathieu \& Merrifield, 2000).  It might therefore reasonably be hoped
that the application of this approach to real kinematic data will
provide a robust method for studying the detailed dynamics of disk
galaxies.

\label{lastpage}

\begin{thebibliography}{99}
\bibitem{b1} Antonov V.A., 1962, Vestnik Leningrad Univ.  19, 96
\bibitem{b2} Bender R., 1990, A\&A 229, 441
\bibitem{b3} Binney J., 1987, in {\it The Galaxy}, eds. G. Gilmore and
B. Carswell (Dordrecht: Reidel), 399
\bibitem{b4} Binney J., Tremaine S., 1987, Galactic Dynamics
(Princeton: Princeton Univ. Press)  
\bibitem{b5} Dejonghe H., Merritt D., 1992, ApJ 391, 531 
\bibitem{b6} Franx M., Illingworth G.D., 1988, ApJ 327, L55
\bibitem{b7} H\'enon M., , 1973, A\&A 24, 229
\bibitem{b8} Kuijken K., Tremaine S., 1991, in {\it Dynamics of Disc 
Galaxies},
ed. B. Sundelius, G\"{o}teborg, 71
\bibitem{b9} Kuijken K., Fisher, D., Merrifield M.R., 1996, MNRAS 
283, 543
\bibitem{b10} Kuijken K., Merrifield M.R., 1993, MNRAS 264, 712
\bibitem{b11} Mathieu A., Merrifield M.R., 2000 (in preparation)
\bibitem{b12} Merrifield M.R., Kent, S.M., 1990, AJ 99, 1548
\bibitem{b13} Merrifield M.R., Kuijken K., 1994, ApJ 432, 575
\bibitem{b14} Merritt D., 1996, AJ 112, 1085
\bibitem{b15} Perez J., Aly J.J., 1996, MNRAS 280, 689 
\bibitem{b16} Pichon C., Thi\'ebaut E., 1998, MNRAS 301, 419 
\bibitem{b17} Rix H.W., White S.D., 1992, MNRAS 254, 389
\bibitem{b18} van der Marel R.P., Franx M., 1993, ApJ 407, 525
\bibitem{b19} Winsall M.L., Freeman K.C., 1993, A\&A 268, 443
\end{thebibliography}
\end{document}